\documentclass[conference]{IEEEtran}
\IEEEoverridecommandlockouts
\usepackage{stfloats}
\usepackage{gensymb}
\usepackage{enumitem}
\usepackage{float}
\usepackage{cite}
\usepackage{tabulary}
\usepackage{amsmath,amssymb,amsfonts}
\usepackage{algorithmic}
\usepackage{graphicx}
\usepackage{textcomp}
\usepackage{xcolor}
\usepackage{booktabs}
\usepackage{wrapfig}
\usepackage{subfigure}
\def\BibTeX{{\rm B\kern-.05em{\sc i\kern-.025em b}\kern-.08em
    T\kern-.1667em\lower.7ex\hbox{E}\kern-.125emX}}
\begin{document}



\title{\LARGE Beyond Point Clouds: A  Knowledge-Aided High Resolution Imaging Radar Deep Detector for Autonomous Driving
}
\vspace{-2mm}
\author{\IEEEauthorblockN{Ruxin~Zheng$^{1}$, Shunqiao~Sun$^{1}$, David~Scharff$^{1}$ and Teresa~Wu$^{2,3}$}  \\
\normalsize $^{1}$Department of Electrical and Computer Engineering, The University of Alabama, Tuscaloosa, AL, 35487\\
$^{2}$School of Computing and Augmented Intelligence, Arizona State University, Tempe, AZ, 85281 \\
$^{3}$ASU-Mayo Center for Innovative Imaging, Arizona State University, Tempe, AZ, 85281
\vspace{-6mm}}

\maketitle

\begin{abstract}
The potentials of automotive radar for autonomous driving have not been fully exploited. We present a  multi-input multi-output (MIMO) radar transmit and receive signal processing chain, a knowledge-aided approach exploiting the radar domain knowledge and signal structure, to generate high resolution radar range-azimuth spectra  for object detection and classification using deep neural networks. To achieve waveform orthogonality among a large number of transmit antennas cascaded by four automotive radar transceivers, we propose a staggered time division multiplexing (TDM) scheme and velocity unfolding algorithm using both Chinese remainder theorem and overlapped array. Field experiments with multi-modal sensors were conducted at The University of Alabama. High resolution radar spectra were obtained and labeled using the camera and LiDAR recordings. Initial experiments show promising performance of object detection using an image-oriented deep neural network with an average precision of $96.1\%$ at an intersection of union (IoU) of typically $0.5$ on $2,000$ radar frames.
\end{abstract}


\vspace{-1mm}
\section{Introduction}

\begin{figure*}[!b]
 \centering
\subfigure[]{\includegraphics[height=2.35 in]{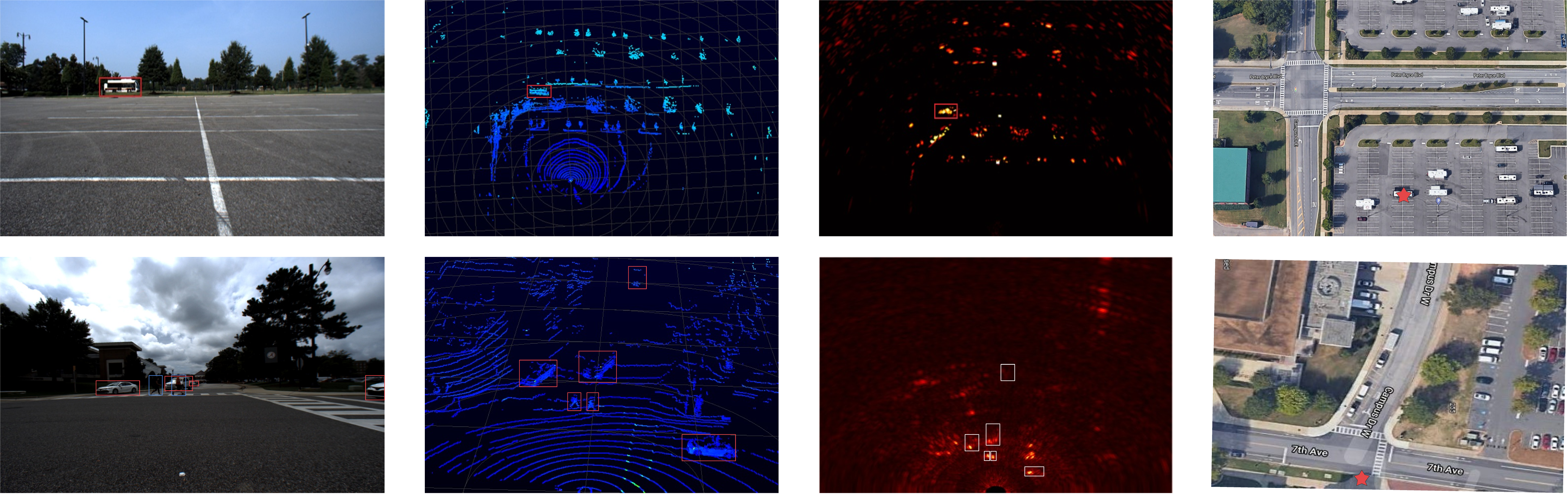}}
\subfigure[]{\includegraphics[height=2.35 in]{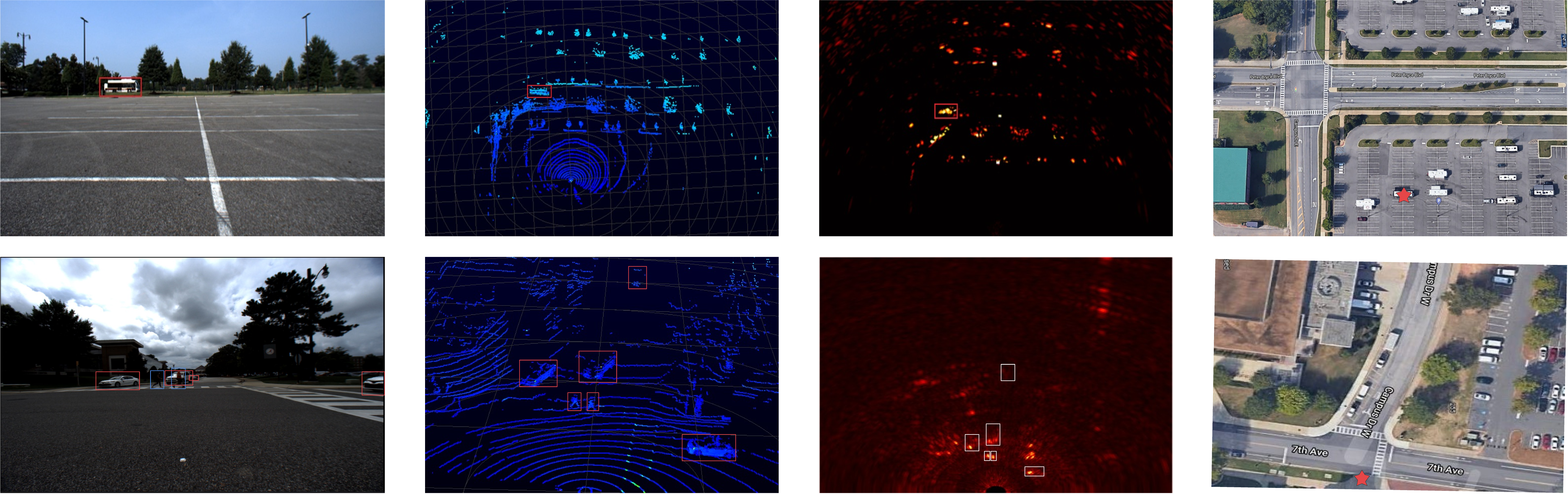}}
\subfigure[]{\includegraphics[height=2.35 in]{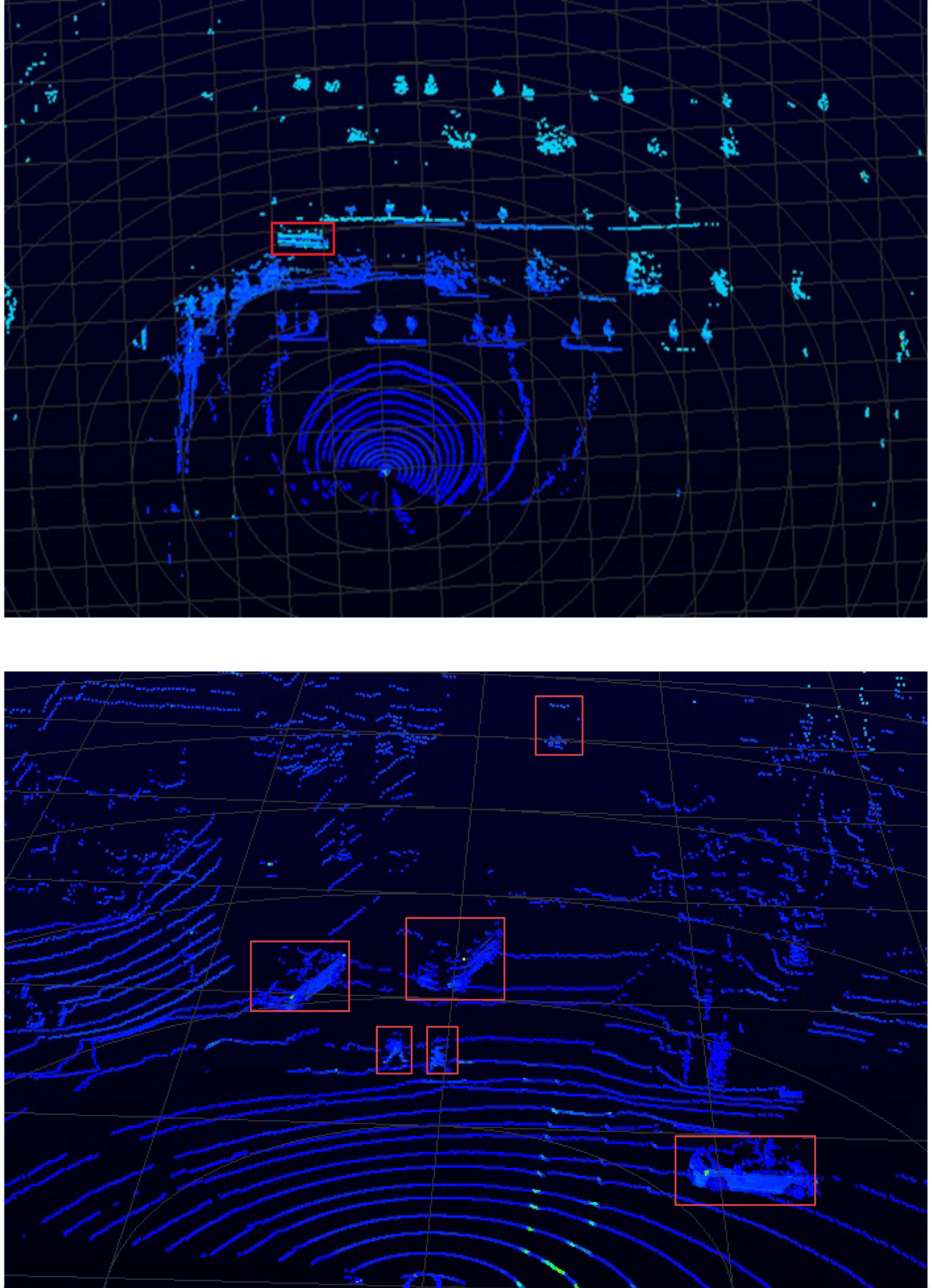}}
\subfigure[]{\includegraphics[height=2.35 in]{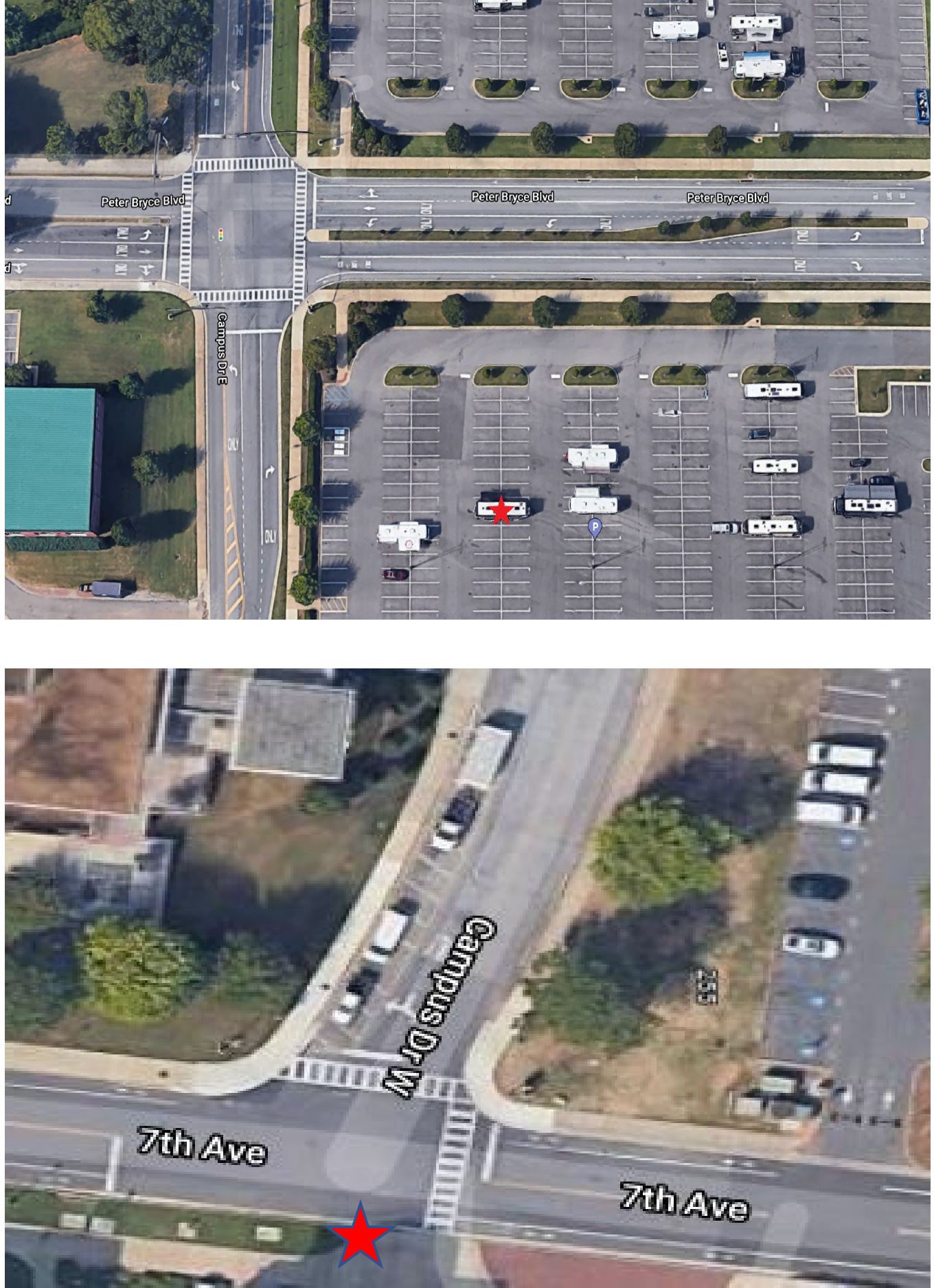}}
\vspace{-2mm}
\caption{Field experiments  with multi-modal sensors  at The University of Alabama:  (a) Camera images, (b): High resolution radar spectra in Cartesian coordinate system, (c): Velodyne Ultra Puck LiDAR 3D point clouds, (d): Google maps  of the field experiment environment, where star \textcolor{red}{$\star$} denotes the sensors location.}
\vspace{0mm}
\label{Radar_data_representation_example}
\end{figure*}

Three types of multi-modal sensors, i.e.,  radar, camera, and LiDAR (light detection and ranging) are widely deployed in autonomous vehicles. In advanced driver assistant system (ADAS), radar sensing is key to enable counter-active measures, such as automatic emergency braking to prevent accidents.
Recently, millimeter wave automotive radar operating at $76$-$81$ GHz has 
emerged as one of the  key technologies in autonomous driving systems, providing  environmental 
perception  under all weather conditions \cite{SUN_SPM_Feature_Article_2020}. Existing automotive radar transceivers, such as NXP Semiconductors MR3003 and Texas Instruments AWR1243 \cite{TI_AWR1243_Transceiver}, support up to $3$ transmit and $4$ receive antennas, yielding angular resolution of around $10^\circ$, which is not capable for Level 4 and Level 5 autonomous driving where a vehicle drives itself in all conditions without any human interaction.  Imaging sensors can detect the traffic lights, road signs and lane markers. However, camera cannot provide accurate depth estimation and its performance degrades greatly under bad weather conditions, or when lighting condition is worse. LiDAR measures the range information of surrounding environment in the form of three dimensional (3D) points with high accuracy. However, LiDAR is quite expensive and its performance degrades significantly under bad weather conditions, such as fog, rain and snow \cite{SUN_SPM_Feature_Article_2020}.

\vspace{-1mm}
High resolution imaging radar systems are highly desired for Level 4 and Level 5 autonomous driving to provide \emph{point clouds} of the surrounding environment  \cite{Bilik_RadarConf_2016,Meinl_ICMIM_2017,Alland_Patent_2018, SUN_SPM_Feature_Article_2020}. Cascaded radar chips rendering $12$ transmit and $16$ receive antennas are being developed \cite{ Bilik_Radarconf_2018,Uhner_PMCW_ISSCC_2019,TI_Cascade} to synthesize a large virtual array using multi-input multi-output (MIMO) radar technology at a low cost \cite{Jian_07,SUN_SPM_Feature_Article_2020}. Several products are available with different array configurations, such as forward-looking full-range radar of ZF and ARS540 of Continental \cite{ZF_Cascade,Continental_Cascade}.  With imaging radar, it is of great interest to investigate environment perception using deep neural networks. The performance of data-driven deep learning depends on the quantity and quality of high dimensional data. If the training data is limited and noisy, the performance deteriorates.

The popular data sets in autonomous vehicle perception, such as KITTI \cite{KITTI} and Waymo Open Dataset \cite{Waymo} only contain camera and LiDAR recordings. Recently, data sets containing automotive radar have been published, such as nuScenes \cite{nuScenes}, Oxford Radar RobotCar \cite{Oxford_radar_robotcar}, Astyx \cite{Astyx}, RADIATE \cite{RADIATE}, CRUW \cite{CRUW}, Zendar 
\cite{Zendar}, CARRADA \cite{CARRADA} and RadarScenes \cite{RadarScenes}. However, most of them are small size and  the angular resolution of automotive radar is low, i.e., larger than $10^\circ$. Some data sets, such as RADIATE, Oxford Radar RobotCar, used a specific radar, such as mechanical scanning radar, which provided denser radar image. However, the Doppler information of targets is missing. Synthetic aperture radar (SAR) technology which is for static targets was adopted in Zendar dataset with multiple measurements from different vehicle locations. The radar angular resolution in CARRADA and CRUW datasets is larger than $10^\circ$. The Astyx dataset is small and only contains sparse radar point clouds.

\begin{table}[h]
\centering
\resizebox{\columnwidth}{!}{%
\begin{tabular}{|l|c|c|c|c|} \hline 
\textbf{Dataset} & \textbf{\# of Frames} &\textbf{Radar} & \textbf{LiDAR} & \textbf{Camera}  \\ \hline
nuScenes \cite{nuScenes}  & $40,000$ & Sparse point clouds & $\surd$ & $\surd$ \\
Oxford Radar  \cite{Oxford_radar_robotcar} & $240,000$  & High resolution radar image & $\surd$ & $\surd$ \\
Astyx \cite{Astyx} & $500$ & Sparse point clouds   & $\surd$ & $\surd$ \\
RADIATE \cite{RADIATE} & $44,000$ & High resolution radar image & $\surd$ & $\surd$ \\
CRUW \cite{CRUW} & $396,241$ & Low resolution radar image  & $ \times$ & $\surd$\\
Zendar \cite{Zendar} & $94,460$ &  SAR with low resolution radar &  $\surd$ & $\surd$ \\
CARRADA \cite{CARRADA} & $12,666$ & Sparse point clouds & $ \times$ & $\surd$\\
RadarScenes \cite{RadarScenes} & $975$ &  Sparse point clouds & $ \times$ & $\surd$
\\ \hline
\end{tabular}}
\vspace{0mm}
\caption{Overview of publicly available radar data sets.}
\label{table_comparison_three_targets}
\vspace{-5mm}

\end{table}

\begin{figure*}
\centering
\includegraphics[width=7.0 in]{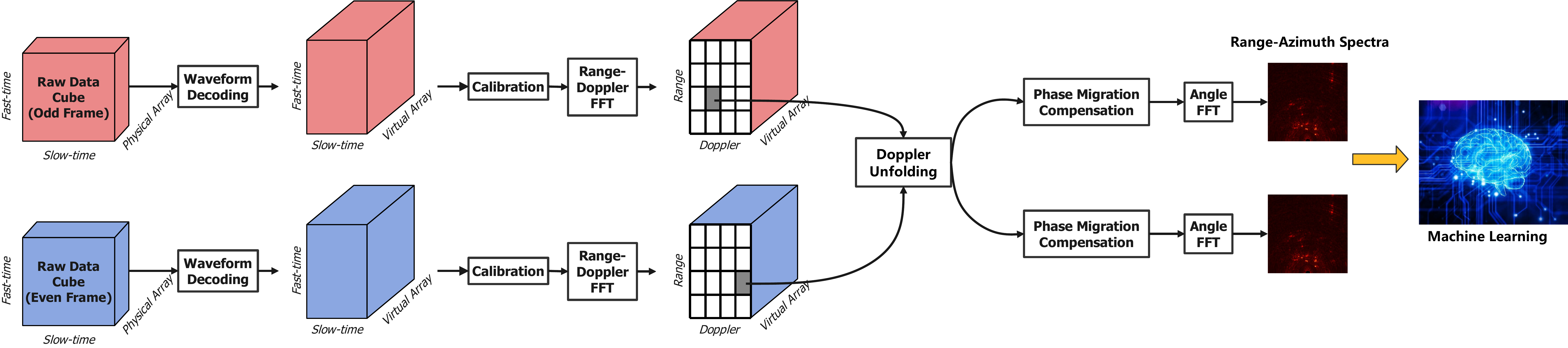}
\vspace{-2mm}
\caption{\label{fig_signal_processing_chain} Proposed radar receive signal processing pipeline to generate high resolution radar spectra for object detection with machine learning.}
\vspace{-5mm}
\end{figure*}

There is information loss in the radar dataset with point clouds, sparse or dense, if thresholding operation has been applied to radar data processing. Domain expert knowledge and signal structure could help enrich the deep learning framework in the scenario of limited training data. In this paper, the goal is to exploit the radar domain knowledge to construct radar data representation without information loss, i.e., high resolution radar spectra that is suitable for the problem space.  Specifically, we aim to optimize the transmit and receive signal processing chain of the imaging radar system to generate high resolution radar range-azimuth spectra representing object's shape. To our best knowledge, there is no open dataset containing high resolution radar spectra  constructed with large number of transmit and receive antennas using the MIMO radar technology.  We configured Texas Instruments (TI) imaging radar \cite{TI_Cascade} with our proposed signal processing chain and conducted field experiments. Two representative examples of radar range-azimuth spectra in Cartesian coordinate system  are shown  in Fig. \ref{Radar_data_representation_example}, where the performance of radar system is close to Velodyne Ultra Puck LiDAR. Based on the high resolution radar spectra, we adopt deep neural networks for object detection and classification.  Our innovations lie in the system level design of MIMO radar transmit and receive signal processing chain, and novel radar data representation that is suitable for object detection using deep learning.

\section{System Model }
In this section, we present a novel radar signal processing chain to generate high resolution radar spectra and object detection approach with deep neural networks.  At transmitter side, we propose a staggered time division multiplexing scheme to synthesize a large virtual array. At receiver side, the radar data is first demodulated and calibrated. We propose a novel velocity unfolding algorithm utilizing both Chinese remainder theory and overlapped array elements. Based on the  estimated unambiguous velocity, the virtual array phase is then compensated.  The proposed radar receive signal processing chain is shown in Fig. \ref{fig_signal_processing_chain}.
\subsection{State-of-the-art FMCW radar }

A FMCW radar transmits a \emph{chirp}, which is a complex sinusoid signal whose frequency changes linearly with time.  The transmit frequency, $f_{T}(t)$, for on chirp with bandwidth $B$ and chirp duration $T$, can be expressed as
\begin{align}
f_{T}\left(t\right) = f_c + \frac{B}{T}t, 
\end{align}
where $f_{c}$ is carry frequency.  
The phase $\varphi_{T}(t)$ of the transmitted signal could be obtained after integration as $\varphi_{T}\left( t \right) = 2\pi\int_{-T/2}^{t}f_{T}\left(t\right) dt$. 
The noiseless received signal is a delay version of transmit signal. For a target at range of $R$ with  radial velocity of $v$, the round-trip delay can be expressed as $\tau= 2(R+v t) /c$.
The received signal is mixed with the transmit signal, and the output of the mixer is called \emph{beat signal}, whose phase could be approximated as 
\begin{align}
\varphi_{B}(t)=
2\pi\left [  
\frac{2f_{c}R}{c}+ \left(\frac{2f_c v}{c}+\frac{2B R}{Tc} \right)t
\right ],
\end{align}
where the beat frequency is $f_b = f_R+f_D$ with $f_R = \frac{2B R}{Tc}$ being the range frequency and $f_D = \frac{2f_c v}{c}$ being the Doppler frequency.
The beat signal typically goes through a band pass filter (BPF) to compensate the gain for targets in distance to improve the radar dynamic range, followed by an analog-to-digital converter (ADC), whose sampling rate is greater than twice of maximum beat frequency $f_b^{\max}$. 
Range and Doppler information of the target could be obtained by applying fast Fourier transforms (FFTs) along fast time and slow time.
\vspace{-0mm}
\subsection{Automotive MIMO  Radar and Waveform Orthogonality}
MIMO radar can synthesize a large virtual array for angle estimation using  multiple transmit and multiple receive antennas \cite{Jian_07,SUN_SPM_Feature_Article_2020}. In MIMO radar, at transmitting side, multiple transmit antennas transmit orthogonal FMCW sequences; at receiving side, due to the waveform orthogonality, the contribution of each transmit antenna can be extracted from the receive signal at each receive antenna. 
There are different ways to achieve waveform orthogonality in MIMO radar, such as Doppler-division multiplexing (DDM) and time division multiplexing (TDM) \cite{SUN_SPM_Feature_Article_2020}. 

In DDM, waveform orthogonality is achieved in Doppler domain by multiplying a binary phase code on each transmitted FMCW chirp. The code is different for each transmit antenna and changes between chirps. The contribution of each transmit antenna can be separated by applying a slow-time Doppler demodulation after range FFT. DDM allows all transmit antennas to transmit simultaneously. However, there is waveform residual from other transmit antennas along Doppler spectrum after Doppler demodulation, which may mask objects with low radar cross section.

\vspace{-3mm}
\begin{figure}[h]
\centering
\includegraphics[width=2.2 in]{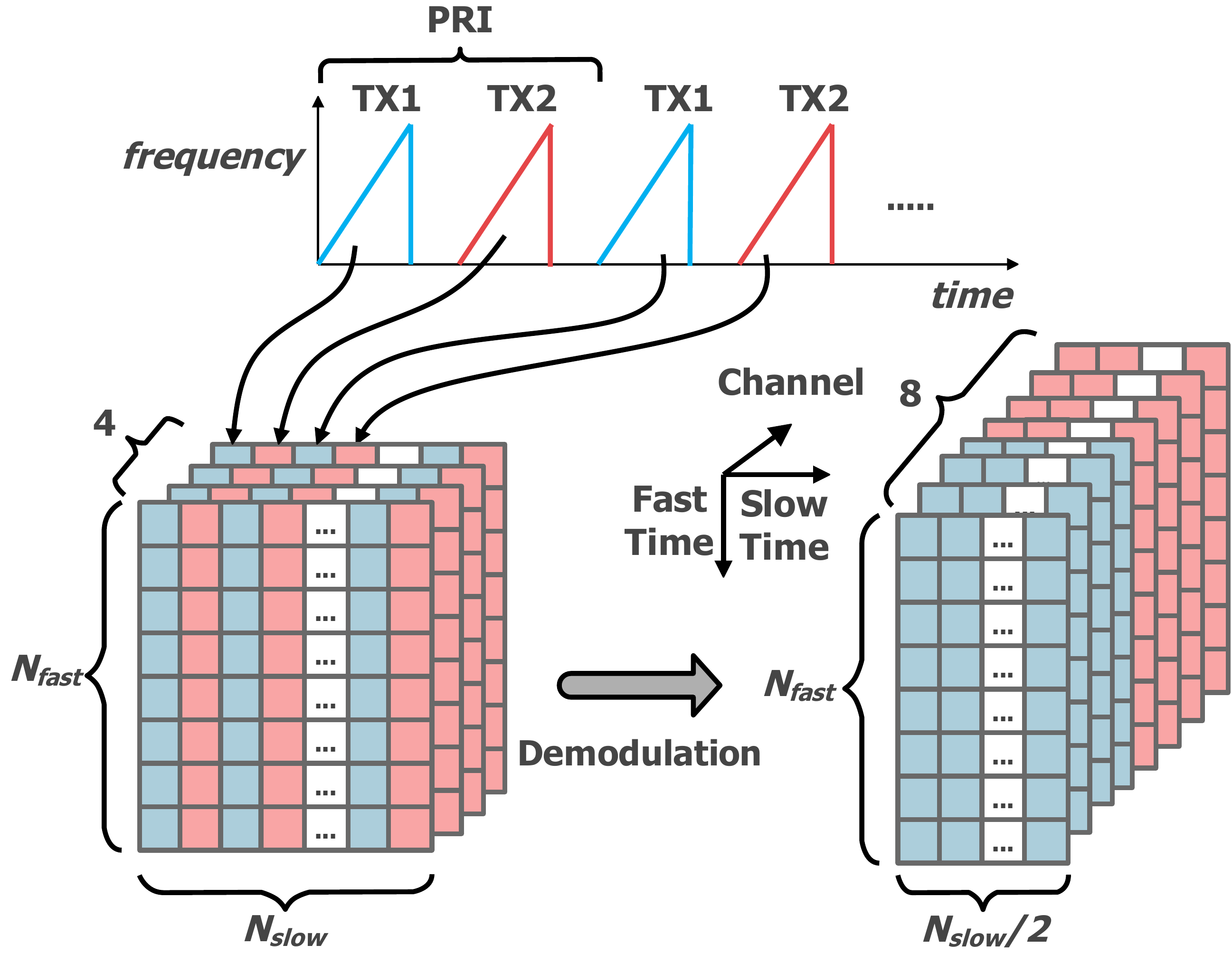}
\caption{\label{fig_TDM} Illustration of waveform orthogonality through TDM.} 
\end{figure}
\begin{figure}[h]
\centering
\subfigure[]{\includegraphics[width=1.70 in]{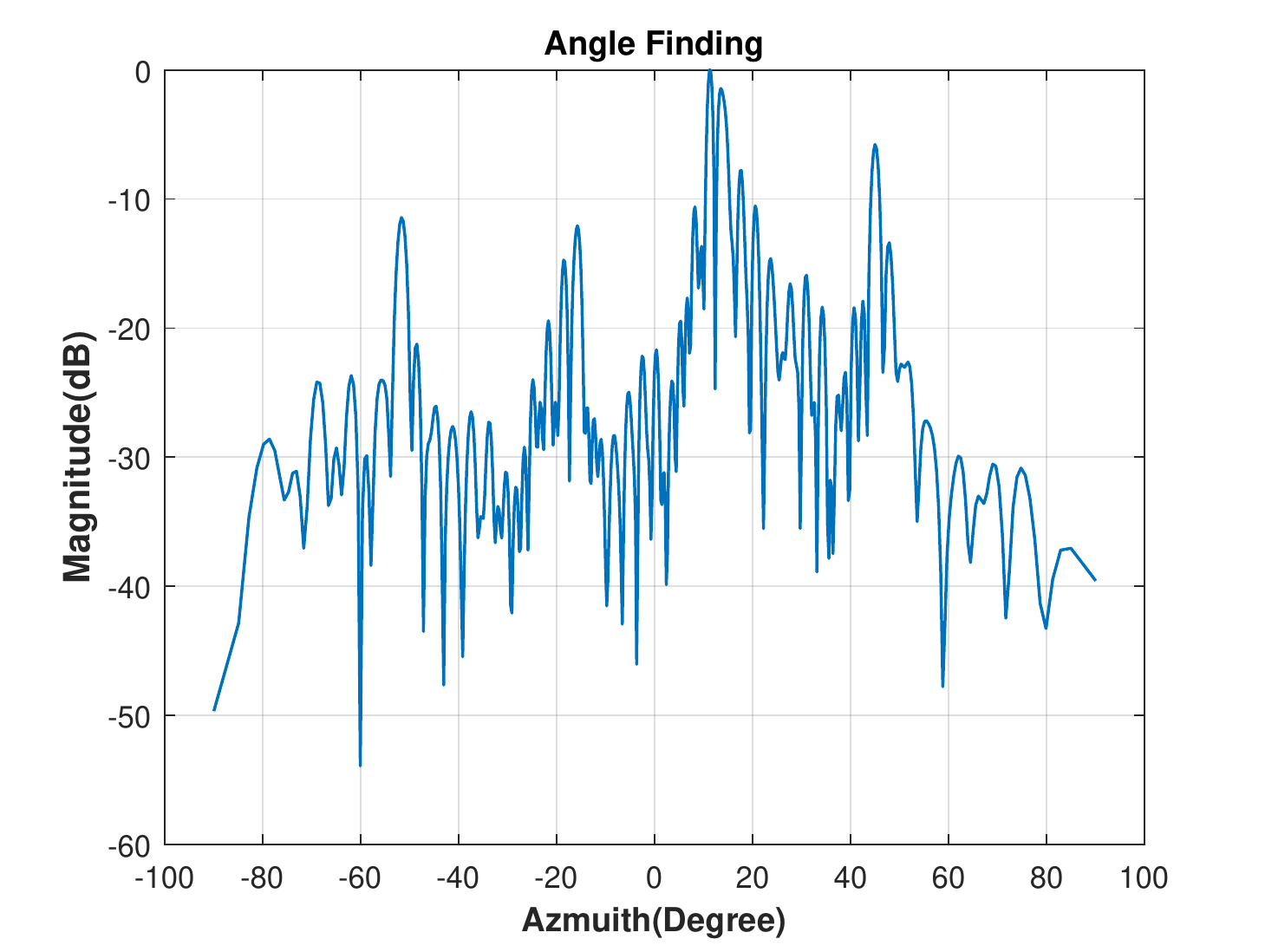}}
\subfigure[]{\includegraphics[width=1.70 in]{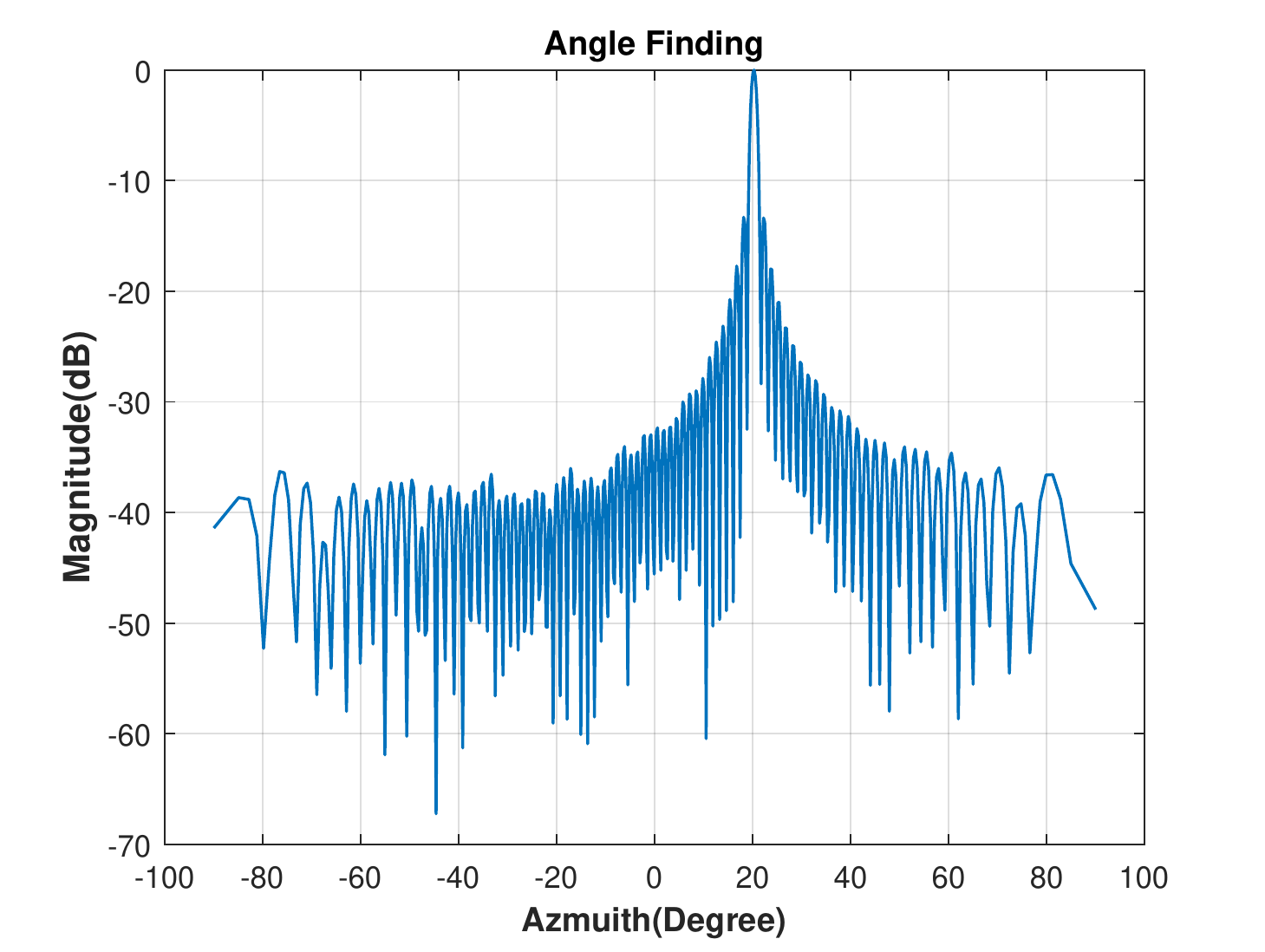}}
\vspace{-3mm}
\caption{\label{fig_angle_spectra_moving_target} Angle spectra of a moving target with velocity of 10 m/s and azimuth angle of $20^\circ$: (a) before and (b) after phase compensation. The radar is configured to select 9 Tx and 16 RX with chirp duration of $50$ $\mu$s.}
\end{figure}

\vspace{-2mm}
Under TDM scheme, only one TX antenna is selected to transmit at each time. A signal processing example of a TDM MIMO radar with two TX and four RX antennas is shown in Fig \ref{fig_TDM}. Assume there are $N_{\rm slow}$ chirps transmitted in one CPI and number of ADC samples is $N_{\rm fast}$. All odd chirps (blue) are  transmitted by TX1; all even chirps (red) are transmitted by TX2. At each receive antenna, 
the radar data matrix can be assembled into two matrices corresponding to odd and even chirp sequences, respectively. Therefore, a radar data cube with dimension of $N_{\rm slow}/2 \times N_{\rm fast} \times 8 $  could be obtained from original $N_{\rm slow} \times N_{\rm fast} \times 4 $ data cube.
The scheduling delay, $\Delta t$, between different transmit antennas would causes phase migration for moving targets between different chirps, i.e.,
\begin{align}
 \phi = (4 \pi / \lambda)v\Delta t.   \label{eq_vel_phase_migration}
\end{align}
That phase migration crates a distortion in virtual array pattern and thus inaccurate angle finding. We show this phenomena via simulation with the same array configuration as TI imaging radar \cite{TI_Cascade}. Assume there is a moving target at $20^\circ$  with $v=10$ m/s. Fig. \ref{fig_angle_spectra_moving_target} (a) plots the angle spectrum obtained from the virtual array without velocity compensation. In order to remove phase migration, for every moving target, a compensation value $e^{-j\phi}$ needs to be multiplied along virtual array before angle finding. 
Fig. \ref{fig_angle_spectra_moving_target} (b) shows the correct angle spectra after compensation. 
\vspace{-3mm}

\vspace{-3mm}

\subsection{Doppler Unfolding by Exploiting the Staggered TDM and Overlapped Arrays}
We adopt TDM to achieve waveform orthogonality due to its simplicity of implementation.
However, under TDM, the maximum unambiguous detectable velocity is reduced to $v_{\max}/N_{\rm TX}$ with $N_{\rm TX}$ being the number of TX antennas \cite{SUN_SPM_Feature_Article_2020}. As $N_{\rm TX}$ increases, the  maximum unambiguous detectable velocity  becomes small and moving targets with relative high speed will be aliased.  We propose a staggered TDM scheme to resolve the Doppler ambiguity using the Chinese remainder theorem (CRT) and overlapped array elements.

\vspace{-3mm}
\begin{figure}[h]%
\centering
{\includegraphics[width=3.3 in]{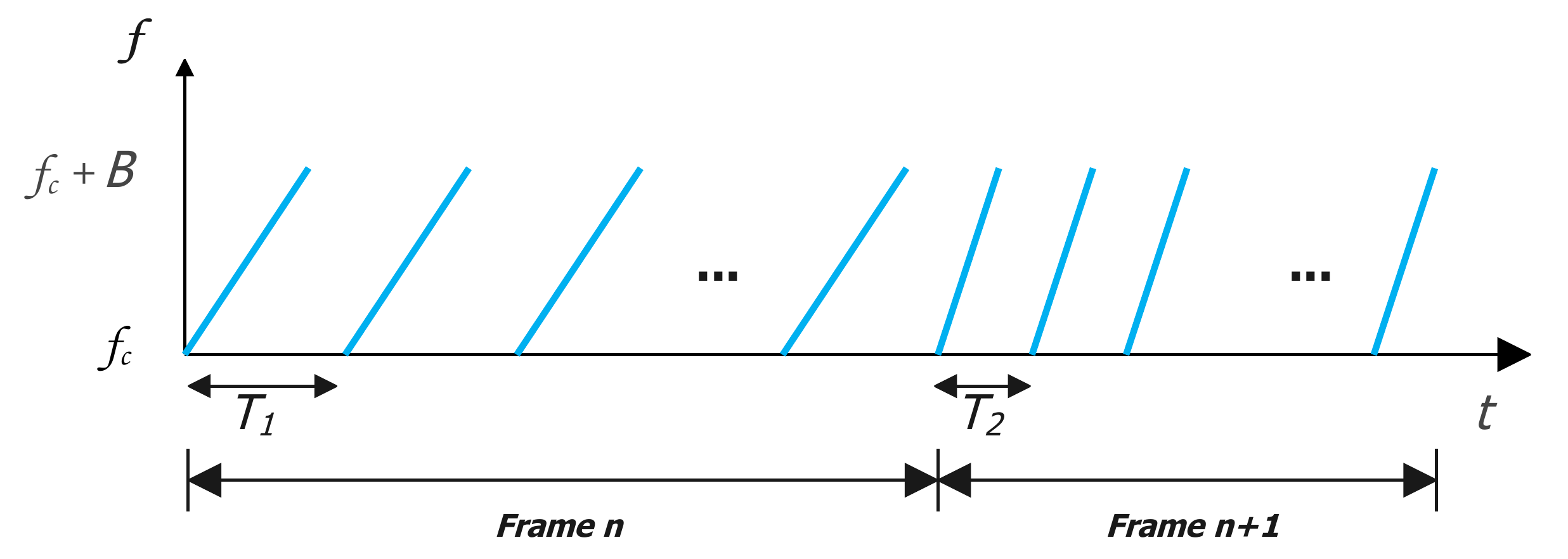}}
\vspace{-2mm}
\caption{Staggered TDM: each transmit antenna is scheduled to transmit  chirps occupying the same bandwidth $B$, every $T_1$ and $T_2$ seconds in consecutive frames. The reduced maximum unambiguous velocity under TDM is then unfolded  by comparing the Doppler detections of the consecutive frames using the Chinese remainder theorem.
}
\label{fig_staggered_TDM}
\vspace{-2mm}
\end{figure}

\subsubsection{Staggered TDM and Chinese Remainder Theory}
Each transmit antenna is scheduled to transmit two consecutive frames with different pulse repetition frequencies (PRFs). In each frame, the chirps occupy the same bandwidth $B$, but with repetition intervals of $T_1$ and $T_2$ respectively, as shown in Fig. \ref{fig_staggered_TDM}. The reduced maximum unambiguous velocity under TDM is then unfolded  by comparing the Doppler detections of the consecutive frames using the Chinese remainder theorem. Different PRFs provide different max unambiguous detectable velocity $v_{\max}$, therefore when the velocity of a moving target exceeds $v_{\max}$ of all frames, the ambiguous velocity estimation $v^m$ of different frames will be different. Assuming the maximal velocity is 90 mph, the set of possible unfolded velocities in the $m$-th frame  could be expressed as 
\begin{align}
{\mathcal S }^m = \left\{ {v^m - 2Mv_{\max }^m, \cdots ,v^m + 2Mv_{\max }^m} \right\}, 
\end{align}
where $v_{\max }^m$ denotes the maximum unambiguous velocity of the $m$-th frame for $m=1,2$, and $M = {{{N_{\rm TX}}} \mathord{\left/
 {\vphantom {{{N_{\rm TX}}} 2}} \right.
 \kern-\nulldelimiterspace} 2}$ if ${{N_{\rm TX}}}$ is even and $M = {{\left( {{N_{\rm TX}} - 1} \right)} \mathord{\left/
 {\vphantom {{\left( {{N_{\rm TX}} - 1} \right)} 2}} \right.
 \kern-\nulldelimiterspace} 2}$ if ${{N_{TX}}}$ is odd. 
The correct velocity can be identified by finding the common velocity candidates in sets ${\mathcal S}^m$ from  consecutive frames. It typically requires four or five different PRFs to resolve velocity ambiguity, which significantly increases the system complexity. We choose to use only two co-prime PRFs, under which a narrow list of possible velocity candidates can be found.
\vspace{-2mm}
\begin{figure}[h]
\centering
\includegraphics[width=2.5 in]{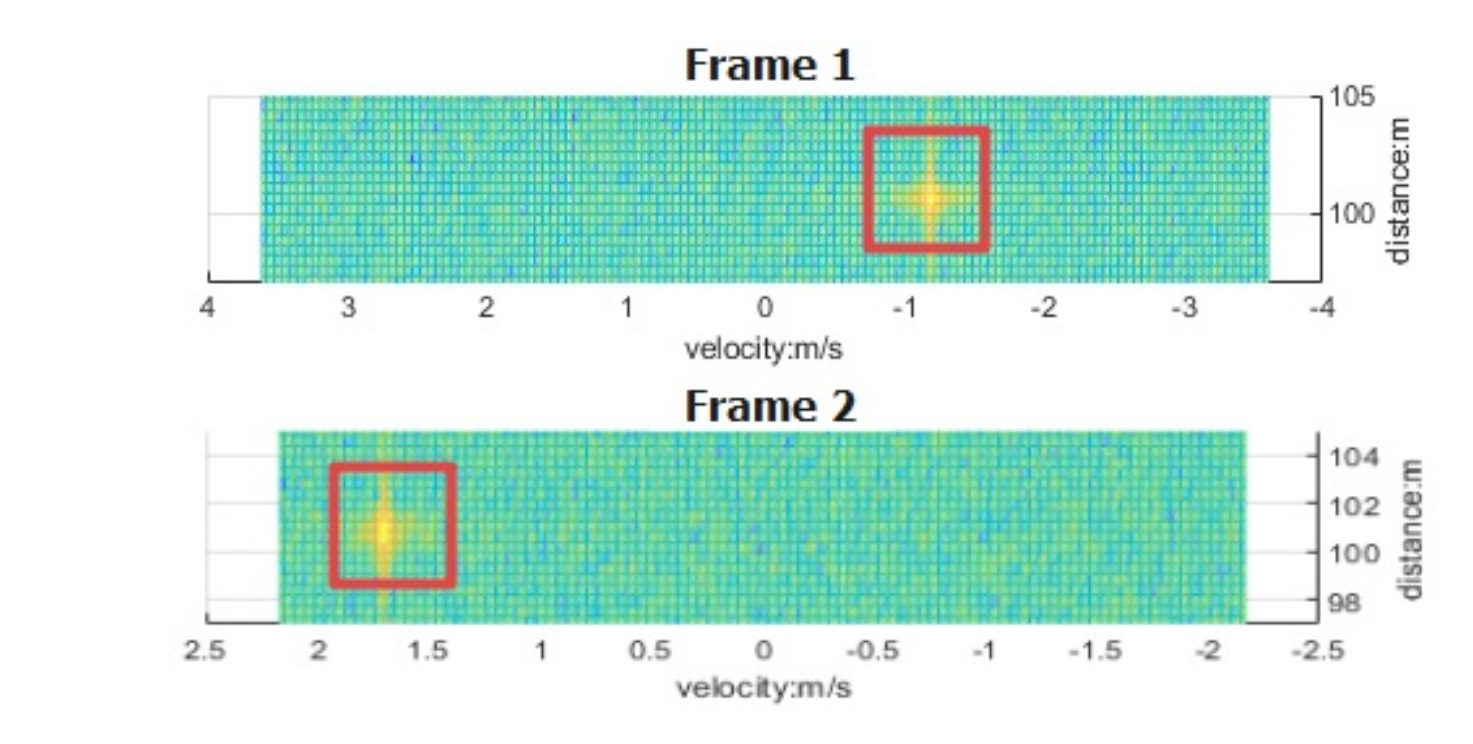}
\vspace{-2mm}
\caption{\label{fig_velocity_ambiguity}  Example of velocity aliasing in the consecutive frames. }
\end{figure}
\vspace{-2mm}

We demonstrate the velocity unfolding with CRT through an example.
Assume a radar is configured to have $v_{\max} = 3.6 $ m/s  on the first frame and $2.2 $ m/s on the second frame and there is a moving target with $6 $ m/s velocity which exceeds $v_{\max}$ of both frames, the target velocity locations on both frames are shown in Fig \ref{fig_velocity_ambiguity}. The detected velocity in frame 1 and fram 2 are respectively  $v^1 =-1.2$ m/s and $v^2=1.7$ m/s. 
The possible unfolded velocities for both frames are respectively $   {\mathcal S}^1 = \left [-30.1, -22.9, -15.6, -8.4, -1.2, 6.0, 13.2, 20.5, 27.7 \right] $ m/s and $    {\mathcal S}^2  = \left[-15.6, -11.3, -7.0, -2.6, 1.7, 6.0, 10.4, 14.7, 19.0 \right]$ m/s.
By finding the common items in sets ${\mathcal S}^1$ and ${\mathcal S}^2$, we can narrow down the actual velocity candidates to $\left[ -15.6, 6.0\right]$ m/s. The final velocity  is determined from the narrow list using overlapped arrays that will be addressed below.

\subsubsection{Overlapped Arrays}
The overlapped array elements corresponding to different transmit antennas have the same azimuth angle phase if the corresponding different TX antennas are transmitting at the same time. 
However, due to the scheduling delay in TDM, the same array element obtained from two different times has phase migration $\phi$, defined in equation (\ref{eq_vel_phase_migration}).  
Therefore, the actual velocity could be obtained by $v = \hat \phi \lambda / (4 \pi\Delta t)$, where $\hat \phi$ is estimated by comparing the phase of overlapped array elements corresponding to different TX antennas. 
This method is easy to implement since it only requires phase comparison for overlapped array elements. However, its detection performance drops significantly when the signal-to-noise ratio (SNR) of overlapped array responses is low. To improve the robustness of velocity unfolding, in this paper, we utilize both CRT and overlapped array.
The narrow velocity candidate list obtained from CRT is used to compensate the phase migration among the overlapped array elements, and the candidate which gives the smallest phase difference will be chosen as actual target velocity.

\section{Radar Data Representation and Deep Neural Networks}

The radar time series data collected from all the receive antennas is a three  dimensional data cube in  fast time, slow time and spatial antenna domain, as shown in Fig. \ref{fig_signal_processing_chain}. Range and Doppler information can be obtained via a two dimensional FFT operation on the radar time series data, 
followed by a constant false alarm rate (CFAR) detector. 
After compensating the data cubes of two consecutive frames, a third FFT is applied across  virtual array elements
to estimate the target's angle information \cite{SUN_SPM_Feature_Article_2020}. The output of automotive radar sensor contains targets' range, velocity, azimuth and elevation angle information, which are referred to as  \emph{point clouds}.
However, a lot of low level radar data representing targets' characters and features were lost if only point clouds are exploited.

\begin{figure}[h]
\centering
\includegraphics[width=3.5 in ]{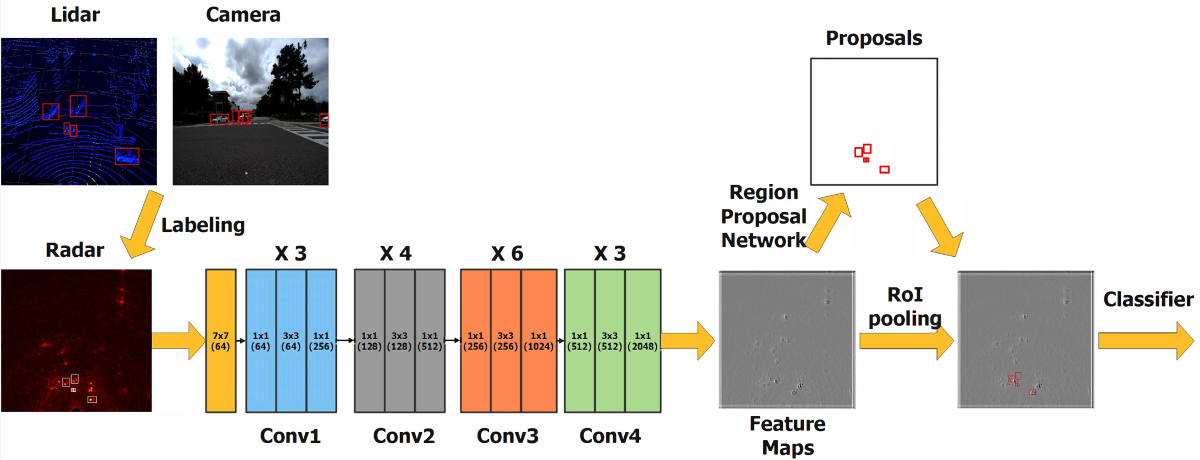}

\caption{\label{fig_radar_ML} Radar machine learning with the Faster R-CNN network.}
\end{figure}
\vspace{-3mm}
Incorporating  structured information into a perception algorithm using machine learning would avoid information loss through the CFAR detector and beamforming of angle finding. In this paper, after the raw radar data passing through the data processing pipeline shown in Fig. \ref{fig_signal_processing_chain}, two range-azimuth spectra are generated, corresponding to odd and even frames. Those spectra are then translated into Cartesian coordinates. The radar spectra generated by FFTs in range, Doppler and spatial domains contain all the information about targets that is available in radar time series. Two representative examples of high resolution bird's-eye-view (BEV) radar range-azimuth angle spectra are shown in Fig. \ref{Radar_data_representation_example}. The high resolution range-azimuth spectra could represent the target shape and radar's performance is comparable to the LiDAR system.

Under supervised learning, a machine learning model is trained on the annotated samples to adjust coefficients of model. Labeling of automotive radar data representation will be a challenging task. Different from images and videos, which are more evident, the data representation of automotive radar is highly abstract, as shown in Fig. \ref{Radar_data_representation_example}. Thus, the manual labeling of automotive radar data requires trained experts. The multi-modal sensors deployed for environmental perception can be utilized to provide transforming data annotations from one sensor to the other. 
In our work, we use the camera images and LiDAR 3D point clouds as ground truth to help manually label the radar data representations. We adopt the Faster R-CNN network \cite{ren2016faster}  with ResNet-50 object detection model for radar machine learning. The radar machine learning pipeline is plotted in Fig. \ref{fig_radar_ML}.

\section{System Evaluation through Field Experiments}
\begin{figure}
\centering
\includegraphics[width=2.2in]{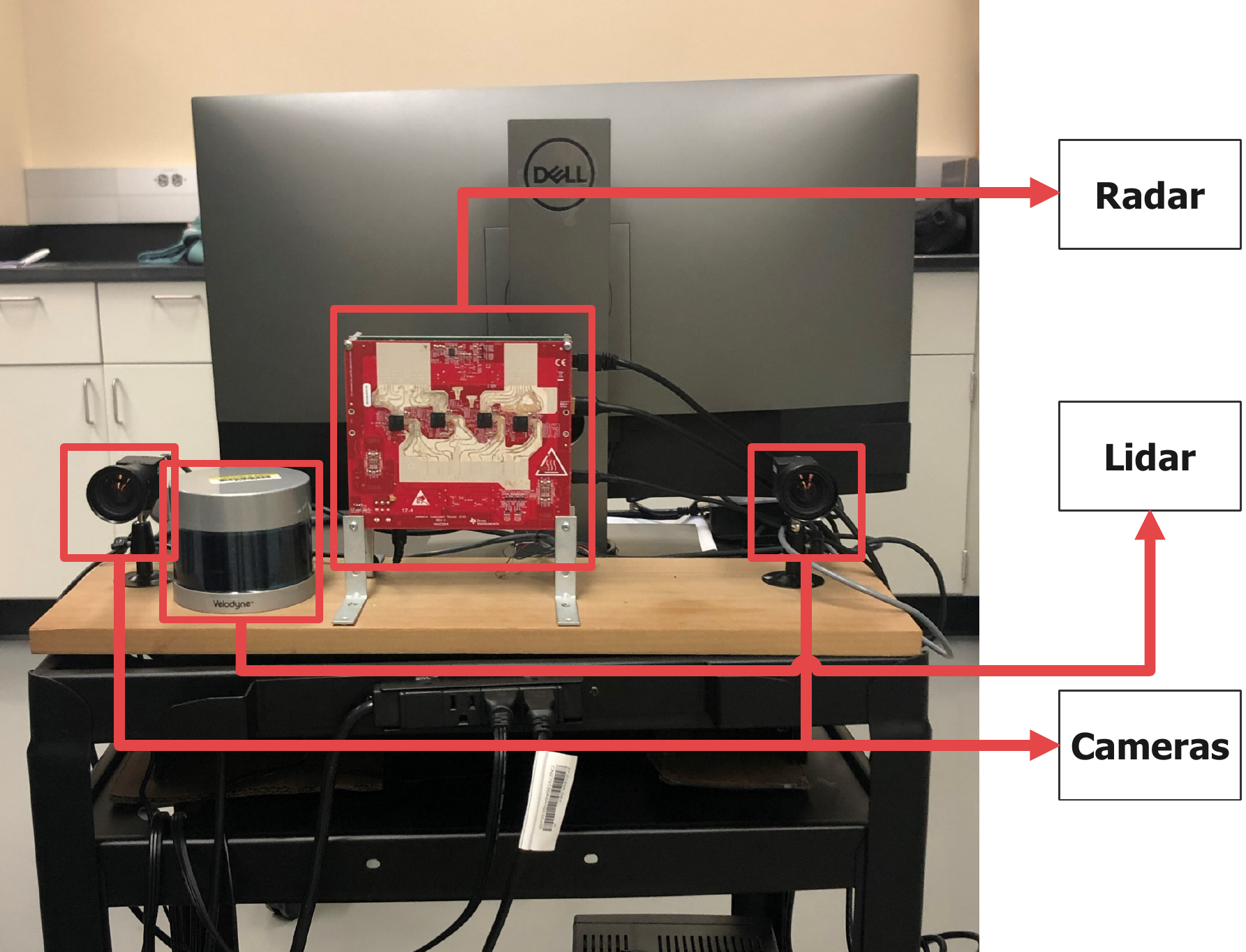}
\caption{\label{fig_multi_modal_sensors} Multi-modal sensors for field experiments. }
\vspace{-6mm}
\end{figure}

Our field experiments included three multi-modal sensors, i.e., a TI imaging radar, stereo cameras of Teledyne FLIR Blackfly S and Velodyne Ultra Puck VLP-32C LiDAR, as shown in Fig. \ref{fig_multi_modal_sensors}. The measurements of cameras and LiDAR are used as ground truth for labeling the radar data. The sensor features are summarized in Table \ref{table_multi_modal_sensors}. 
\begin{table}[h]
\centering
\resizebox{\columnwidth}{!}{%
\begin{tabular}{|l|l |c|} 
 \hline
{\bf Sensors} & {\bf  Model}\\  
 \hline
 Radar & TI Imaging Radar, Azimuth Resolution: $1.2^\circ$, Azimuth FOV: $70^\circ$ \\ 
 \hline
 LiDAR &  Velodyne Ultra Puck VLP-32C, Azimuth Resolution: $0.1^\circ$ - $0.4^\circ$\\& Vertical FOV: $40^\circ$, Maximum Range: 200 m \\
 \hline
 Camera & Teledyne FLIR Blackfly S, Stereo, Image Resolution: 2048 × 1536 \\
 \hline
\end{tabular}
}
\vspace{0mm}
\caption{Multi-Modal Sensors }
\label{table_multi_modal_sensors}
\vspace{-4mm}
\end{table}
\vspace{-3mm}

TI imaging radar is a chirp configurable MIMO radar, which has $12$ TX  and $16$ RX antennas, cascaded by $4$ radar transceivers. The azimuth field of view (FOV) is $70^\circ$. A virtual uniform linear array with $86$ elements and half wavelength spacing can be synthesized with $9$ TX and $16$ RX antennas, of which $32$ virtual array elements are overlapped.  The 3 dB beam width of the imaging radar in azimuth is $    \Delta\theta_{\rm AZ} = 2 {\rm arcsin}\left ( \frac{1.4\lambda}{\pi D_x} \right ) \approx 1.2^\circ$, where $D_x = 42.5\lambda$ is the virtual array aperture in horizontal direction. 
The $9$ TX antennas are scheduled to transmit FMCW chirp sequence in a TDM fashion. Two consecutive frames with different PRFs and overlapped arrays are utilized to resolve the ambiguous velocity problem. The parameters of consecutive frames are shown in Table \ref{table_radar_parameters}. Antenna calibration is required to reduce the frequency, phase and amplitude mismatches across those $4$ radar transceivers. A one-time boresight calibration method is used as our calibration method.
To generate a calibration matrix, a corner reflector is placed at range of $5$ m along the boresight direction  (see Fig. \ref{fig_experiment_cali_car} (a)).
\begin{table}[h]
\centering
\begin{tabular}{|l |l|c|} 
 \hline
 {\bf Parameters} & {\bf Frame $n$} & {\bf Frame $n+1$}\\
 \hline
 Max Range & 150 m & 150 m\\ 
 \hline
 Max Velocity & 5.15 m/s & 3.97 m/s\\ 
 \hline
 Range Resolution & 0.6 m & 0.6 m\\
 \hline
 Angle Resolution & $1.2^\circ$  & $1.2^\circ$ \\ 
 \hline
\end{tabular}
\vspace{2mm}
\caption{Radar Parameters}
\label{table_radar_parameters}
\vspace{-5mm}
\end{table}

\subsection{Doppler Unfolding for Moving Targets}
To validate the proposed velocity unfolding method, we carried out a couple of experiments with one moving car in an open parking lot. Fig. \ref{fig_experiment_cali_car} (b) shows a car moving towards the radar at a constant speed of $10$ mph along the approximate boresight direction. Fig. \ref{fig_experiment_cali_car} (e)  indicates that the consecutive frames yield different Doppler estimation indices under the staggered TDM. The velocity unfolding and phase compensation are then carried out. It can be found in Fig. \ref{fig_experiment_cali_car} (f) that without phase compensation, the angle of moving car is wrongly estimated as $-5.5^\circ$, while with phase compensation, the angle of moving car is accurately estimated as $1.2^\circ$.   

\vspace{-3mm}

\subsection{Range and Angle Resolution Validations}
\begin{figure*}
\centering
\subfigure[]{\includegraphics[height=1.32 in]{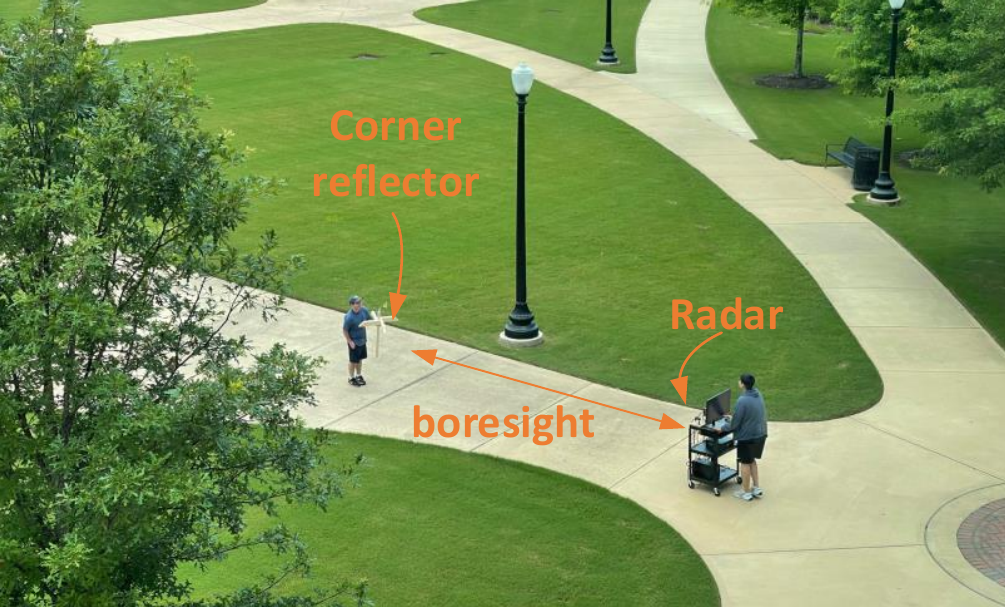}}
\subfigure[]{\includegraphics[height=1.32 in]{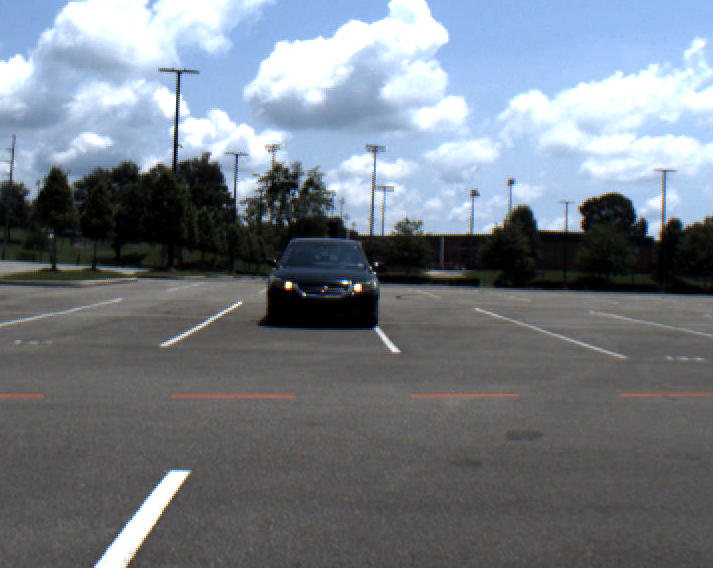}}
\subfigure[]{\includegraphics[height=1.32 in]{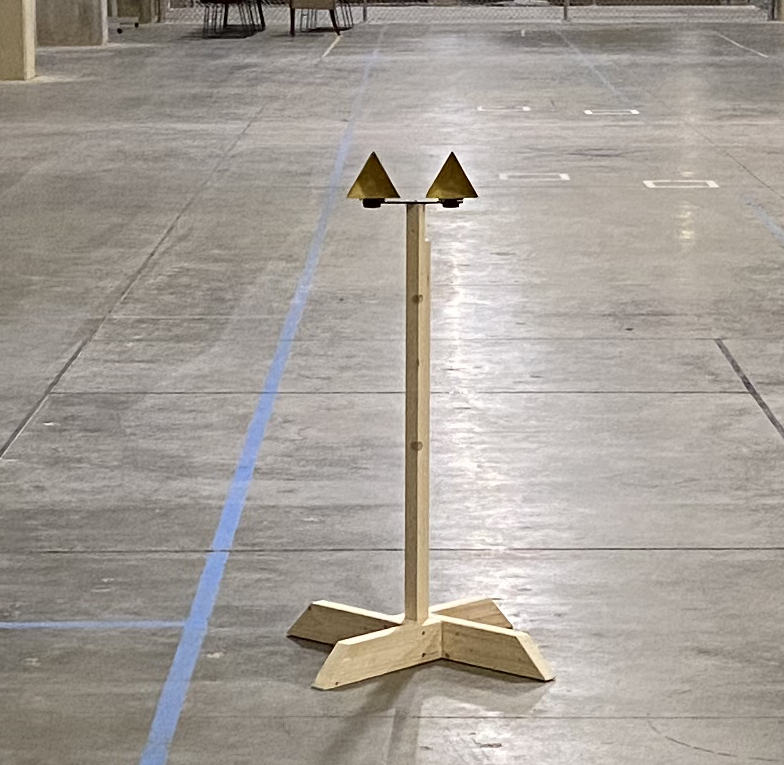}}
\subfigure[]{\includegraphics[height=1.32 in]{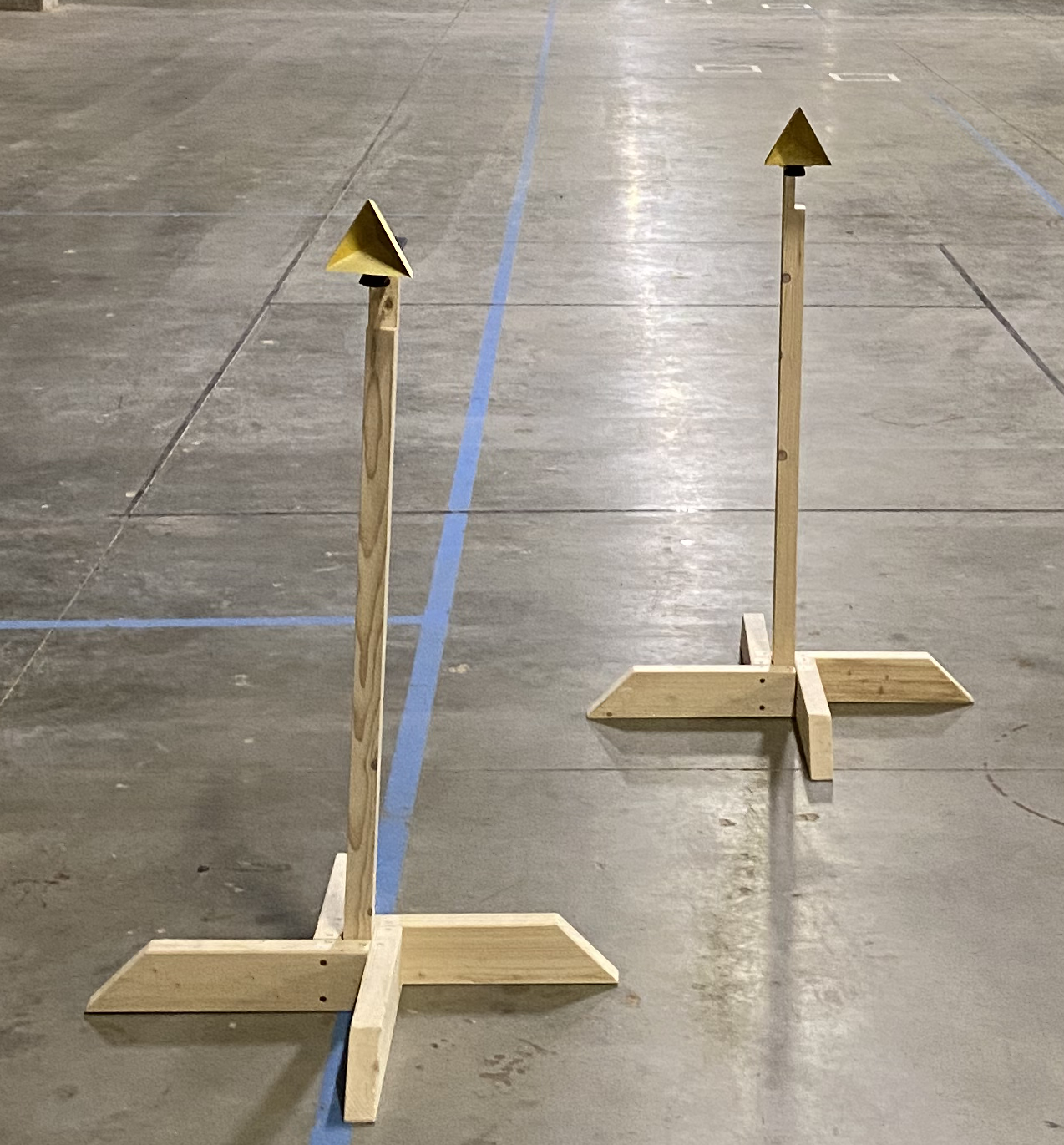}}
\subfigure[]{\includegraphics[height= 1.0 in]{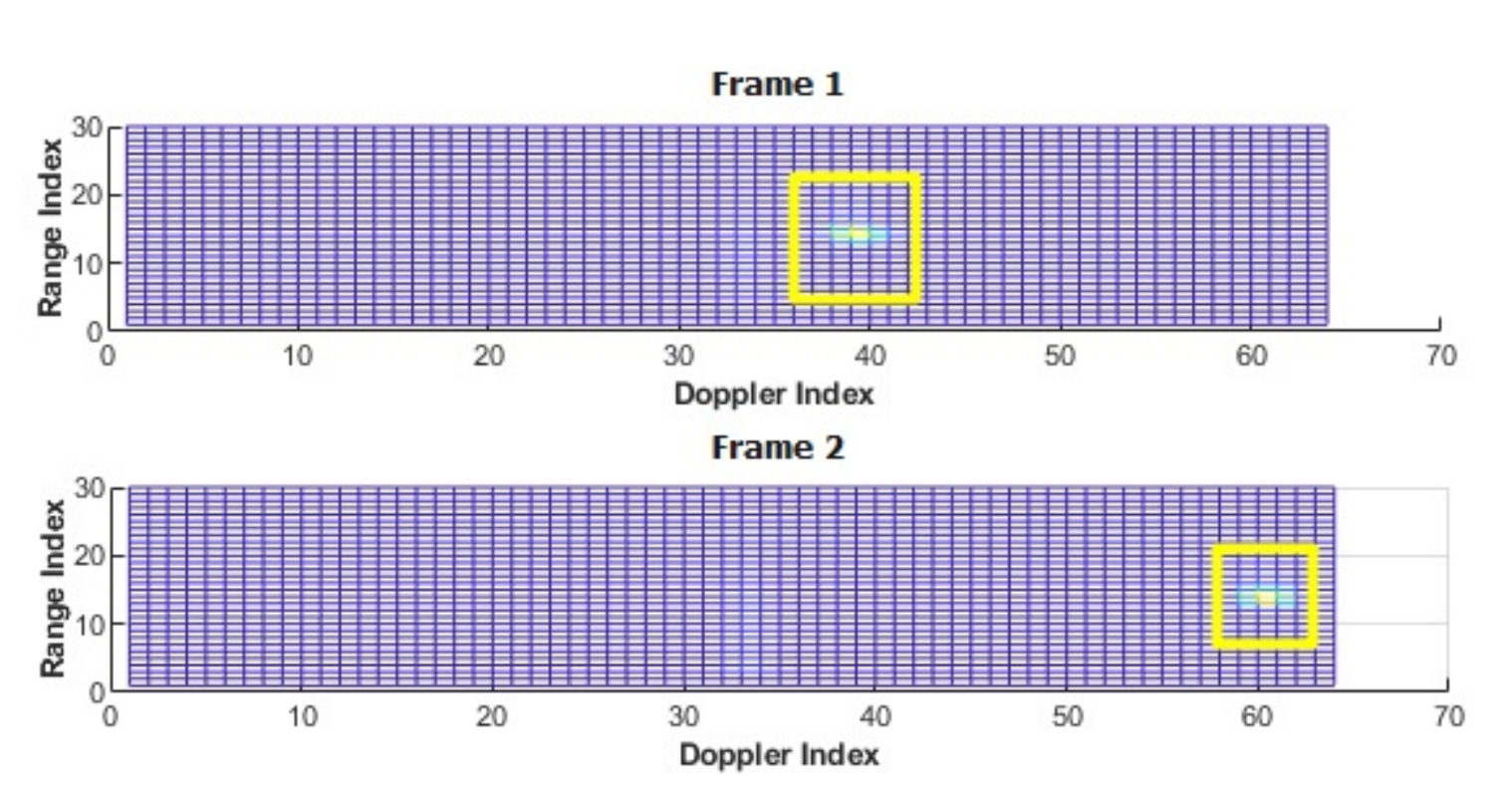}}
\subfigure[]{\includegraphics[height=1.2 in]{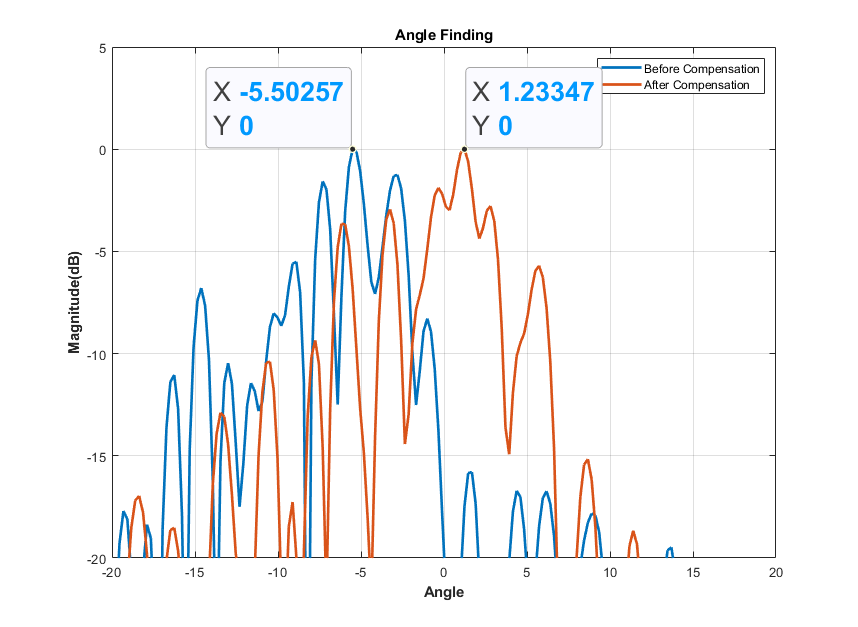}}
\subfigure[]{\includegraphics[height=1.2 in]{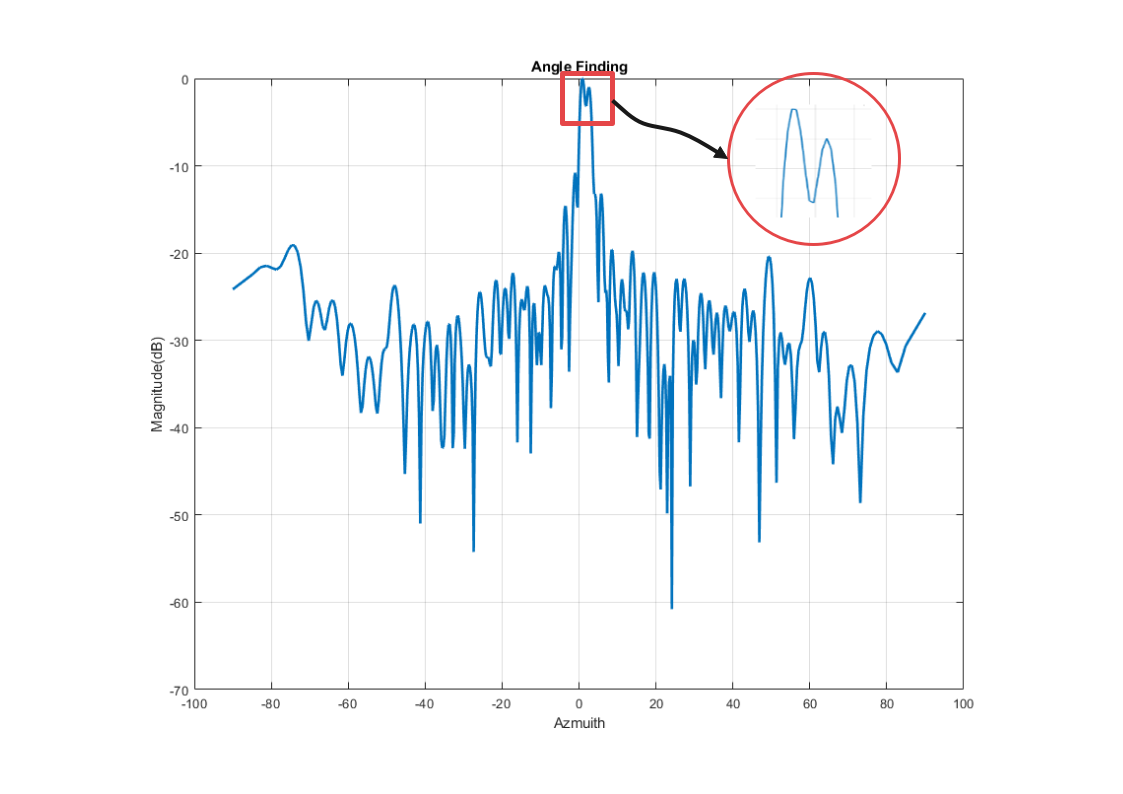}}
\subfigure[]{\includegraphics[height=1.2 in]{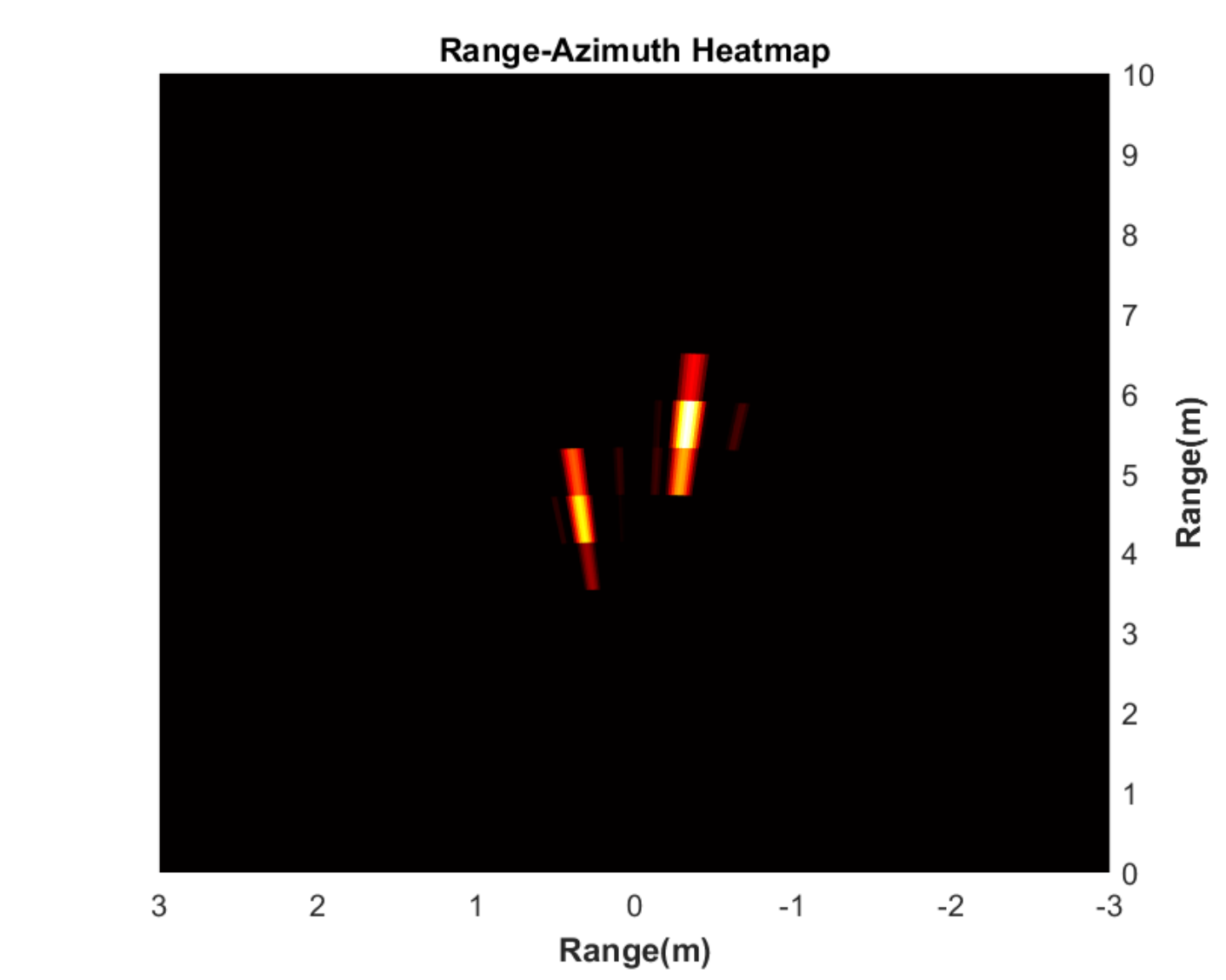}}
\caption{\label{fig_experiment_cali_car} Field experiments. (a) radar calibration, (b) a car is moving at speed of $10$ mph towards the radar, (c) two corner reflectors are separated by $1.6^\circ$, (d) two corner reflectors are separated by $0.6$ m, (f) angle spectrum before and after phase compensation, (g) angle spectrum, (h) range-azimuth spectrum.}
\vspace{-5mm}
\end{figure*}

We validated the range and angle resolution of our radar configuration. Fig. \ref{fig_experiment_cali_car} (c) shows our angle resolution experiment set-up, where two corner reflectors are placed at range of $5$ m along the  boresight direction. Those two corner reflectors are $14$ cm apart, corresponding to $1.6^\circ$ separation. The angle spectrum plotted in Fig. \ref{fig_experiment_cali_car} (g)  shows the imaging radar is capable to distinguish those two corner reflectors. 
In the second experiment, two corner reflectors are placed in front of the radar with separated distance of $0.6$ meters (see Fig. \ref{fig_experiment_cali_car} (d)). The range azimuth spectrum in Cartesian coordinates is shown in Fig. \ref{fig_experiment_cali_car} (h), where it can be found that two corner reflectors can be successfully separated.  

\subsection{Evaluation of Target Detection with Deep Neural Networks}
Our data was collected at an intersection on The University of Alabama campus, shown in Fig. \ref{Radar_data_representation_example} (a), for $15$ minutes which contains $2,000$ radar frames with two object categories, i.e., pedestrian and cars. Those objects in radar frames were labeled manually using LiDAR and camera data as ground truth and randomly divided into a  training set containing $1,200$ frames and a  testing set containing $800$ frames. We adopted a Faster R-CNN network \cite{ren2016faster} with ResNet-50 object detection model, pre-trained on COCO 2017 dataset \cite{lin2014microsoft}  with training images scaled to $640\times640$, as our neural networks. The Faster R-CNN neural networks was trained on a Lambda Lab workstation with an Intel i9 CPU and dual Nvidia RTX 2080 Ti GPUs. Table \ref{table_radar_ML} summarizes the radar machine learning performance. 

\begin{table}[h]
\centering
\begin{tabular}{|l| c |c| c| c|} 
 \hline
 {\bf Model} &  {\bf AP} & {\bf AP@0.5} & {\bf AP@0.75} & {\bf AR}\\
 \hline
 Faster R-CNN Resnet-50  & 75.6\% & 96.1\% & 89 \%& 79.7\%\\ 
 \hline
\end{tabular}
\vspace{2mm}
\caption{Evaluation Metrics}
\label{table_radar_ML}
\vspace{-5mm}
\end{table}

We adopted the evaluation metrics of COCO dataset to evaluate the radar machine learning performance. As shown in Table \ref{table_radar_ML}, the average precision (AP) is $75.6\%$; AP of intersection of union (IoU) corresponding to $0.5$ is $96.1\%$, and AP of IoU corresponding to $0.75$ is $89\%$. Since the detected targets in our dataset are all under size of $32\times 32$, the average recall (AR) of small targets is used to rate our model, which is $79.7\%$. We show that an image-oriented neural network has a very promising 
performance on high resolution radar spectra for vehicle and pedestrian detection and classification, even when the radar dataset size is small.

\section{Conclusions}
In this paper, a staggered TDM scheme and signal processing chain were proposed to achieve waveform orthogonality among a large number of transmit antennas to synthesize a large virtual antenna array to generate high resolution radar range-azimuth spectra. We unfolded the ambiguous velocities and compensated the phase migration among antenna array due to scheduling delay under TDM.  Via field experiments, we demonstrated the promising performance of object detection and classification using deep neural networks on  high resolution radar range-azimuth spectra, that unlocks the radar potentials for Level 4 and Level 5 autonomous driving.

\bibliographystyle{IEEEtran}
\bibliography{refs}

\end{document}